\documentstyle[rotate,psfig]{mn}
\newcommand{\bugref}{\bibitem[\protect\citename{dummy }1893]{dum}}

\begin{document}

\title[Unusual Radio Properties of the BL Lac Object 0820+225]
{Unusual Radio Properties of the BL Lac Object 0820+225} 

\author[D.~C.~Gabuzda, A.~B.~Pushkarev \& N.~N.~Garnich]
{D.~C.~Gabuzda$^{1,2}$, A.~B.~Pushkarev$^2$ and N.~N.~Garnich$^3$\\
$^1$Joint Institute for VLBI in Europe, Postbus 2, 7990 AA Dwingeloo,
The Netherlands\\
$^2$Astro Space Centre, Lebedev Physical Institute, 53 Leninsky Pr., 
117924 Moscow, Russia\\
$^3$Sternberg Astronomical Institute, 13 Universitetskii pr.,
119899 Moscow, Russia\\
}

\maketitle
\begin{abstract}
We present the results of simultaneous VLBA polarisation observations
of the BL Lacertae object 0820+225 at 5, 8, and 15~GHz, together with
earlier images at 5~GHz. This source has
an unusually rich total intensity and polarisation structure compared
to other objects with comparable redshifts.  The magnetic field in the
inner part of the complex and highly twisted VLBI jet is transverse,
as is typical of BL Lacertae objects, but becomes roughly longitudinal
further from the core, possibly due to shear. Although the integrated
rotation measure of 0820+225 is modest, the rotation-measure distribution
on parsec scales is non-uniform, and clearly shows regions where the
rotation measure is substantially higher than the integrated value.
\end{abstract}
\section{Introduction}
BL Lacertae objects are highly variable, polarized, flat spectrum AGN that
are distinguished from OVV quasars primarily by the absence of strong optical
line emission.  5, 8.4, and 22~GHz global VLBI polarization observations
have shown that the jet components in BL Lacertae objects are most often 
polarized with electric field parallel to the jet (Gabuzda, Pushkarev, \&
Cawthorne 2000; Gabuzda \& Cawthorne 1996, 2000; and references therein).
This has often been
interpreted as evidence for transverse shocks that have compressed an
initially random magnetic field {\bf B}, so that the net {\bf B} in
the shocked region is perpendicular to the jet (Laing 1980; Hughes, Aller,
and Aller 1989). Recently, evidence has been mounting that many of the
transverse {\bf B} fields observed in the VLBI jets of these sources
may instead be associated with the dominant toroidal component of a
possibly helical jet {\bf B} field (Gabuzda 1999; Pushkarev \& Gabuzda
2000; Gabuzda \& Pushkarev 2001). We are undertaking a
multi-frequency, multi-epoch VLBI study of the complete sample of
BL Lacertae objects defined by K\"uhr and Schmidt (1990) in order to
uncover the nature of the characteristic VLBI properties of these sources.

One of the advantages of making systematic observations of
a complete sample is that we obtain information about a wide range
of sources, not only a few of the brightest and, at first glance, most
spectacular and interesting. This has proved to be very valuable in our
multi-frequency (5, 8.4 and 15~GHz; February and April 1997) VLBA 
observations of sources in the
K\"uhr \& Schmidt sample. The object 0820+225 had not been studied
previously using VLBI.  This source has proved itself to be
a rather unusual member of this sample for a number of reasons.
Its integrated 5-GHz flux is of the order of 1 Jy, but its VLBI
structure is very extended, and the VLBI core contains only a
small fraction of the total mas-scale flux. This is particularly
remarkable because 0820+225 is one
of only a half dozen of the sample sources with large redshifts,
$z\sim 1$ or slightly greater. This combination of extremely
extended structure at high redshift makes it unique among the sample
sources. First-epoch 5~GHz images (Gabuzda, Pushkarev,
\& Cawthorne 2000) show that the VLBI jet is highly curved,
and forms an overall ``S-like'' shape. The origin of this curvature is
unclear; it is suggestive of some type of instability in or precession
of the VLBI jet.

The two 5~GHz images presented here are second and third-epoch
images. Overall, the 5~GHz VLBI structure of the source is relatively
stable, showing a number of features that appear to be nearly
stationary.
The multi-frequency images presented here reveal the presence of
substantial rotation-measure (RM) gradients both along and across the jet.
After accounting for the parsec-scale RM distribution, the polarization 
vectors in the inner section of the jet are well aligned with
the local jet direction, so that the magnetic field in this part of
the jet is transverse. The polarization in the more extended part
of the jet is somewhat offset toward the outer edge of the jet and
implies a longitudinal field, suggesting a shear interaction
with a surrounding medium. The observed rotation measures suggest that
the thermal gas causing the Faraday rotation has properties that are
fairly typical of narrow-line clouds. 

\section{Observations and Reduction}

\subsection{5~GHz, epoch 1995.36}

First-epoch global-VLBI $I$ and $P$ images of 0820+225 at 5~GHz have
been presented by Gabuzda, Pushkarev, \& Cawthorne (2000).  Our second-epoch
observations were obtained at epoch 1995.36 using a ten-station global
array consisting of the Medicina (32~m), Effelsberg (100~m),
St. Croix (25~m), Hancock (25~m), Green Bank (43~m), North Liberty (25~m),
Owens Valley (25~m), Brewster (25~m), and Mauna Kea (25~m) antennas and
the Westerbork phased array ($\sqrt{14}\times 25$~m).
The observations were made under the auspices of the US and European VLBI
networks. The north--south
resolution provided by these observations is substantially improved
due to the use of the two southern antennas St. Croix and Mauna Kea.
The data were recorded using the MkIII system, and the data were
subsequently correlated using the Mk~IIIA correlator at
the Max-Planck-Institut-f\"ur-Radioastronomie in Bonn.

The data reduction and imaging were done in the Brandeis VLBI package.
The polarization calibration of these data was performed as described by
Roberts, Wardle and Brown (1994). The instrumental polarizations for
each antenna were determined from observations of the unpolarized
sources 3C84 and OQ208. The absolute orientation of the polarization
position angles was calibrated using VLBI observations of several
compact polarized sources together with measurements of their integrated
polarizations derived from nearly simultaneous VLA observations. The
final uncertainty in the polarization position angles is about $2-3^{\circ}$.

Images of the distribution of total
intensity $I$ were made using a self-calibration algorithm similar to that
described by Cornwell and Wilkinson (1981). Maps of the linear
polarization\footnote{$P = pe^{2i\chi} = mIe^{2i\chi}$, where $p=mI$ is the
polarized intensity, $m$ is the fractional linear polarization, and $\chi$ is
the position angle of the electric vector on the sky, measured from north
through east.} $P$ were made by referencing the calibrated cross-hand
fringes to the parallel-hand fringes using the antenna gains determined
in the hybrid mapping, Fourier transforming the cross-hand fringes,
and performing a complex CLEAN. This procedure registers the
$I$ and $P$ maps to within a small fraction of a beamwidth.

\subsection{Multi-frequency, epoch 1997.11}

Our multi-frequency observations of 0820+225 were made
in February 1997 (1997.11)
with the ten antennas of the NRAO Very Long Baseline Array\footnote{The
National Radio Astronomy Observatory (NRAO) is a facility of the National
Science Foundation operated under cooperative agreement by Associated
Universities, Inc.}. Observations were obtained simultaneously
at 15, 8.4, and 5~GHz with full polarization sensitivity. The data were
calibrated and imaged in the NRAO AIPS package using standard techniques.

The strong source 3C84, which is essentially unpolarized at these
wavelengths, was used to determine the instrumental polarizations of
each antenna at each frequency. The absolute orientation of the
polarization position angles was calibrated using VLBA observations
of the compact, strongly polarised source 1823+568 together with
measurements of its integrated polarization from VLA data obtained
roughly one day after the end of the VLBA run. Here, also, the resulting
uncertainty in the absolute $\chi$ values is about $2-3^{\circ}$ at
all frequencies.

\section{Results and Discussion}

\begin{figure}
\hspace*{-0.7cm}
\mbox{
\rotate[r]{\psfig{file=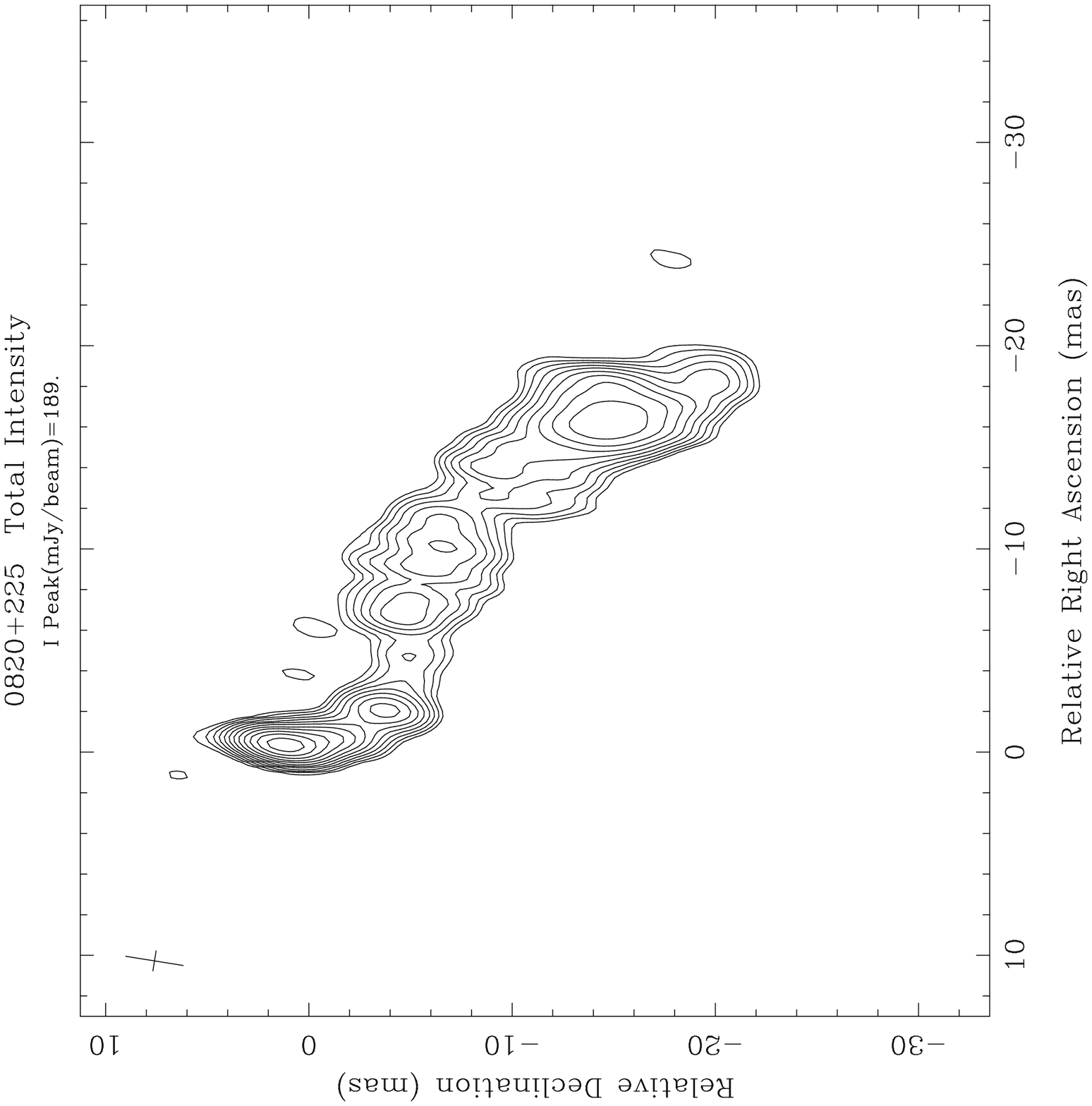,height=10.25cm}}
}
\vspace*{-0.75cm}

\hspace*{-0.7cm}
\mbox{
\rotate[r]{\psfig{file=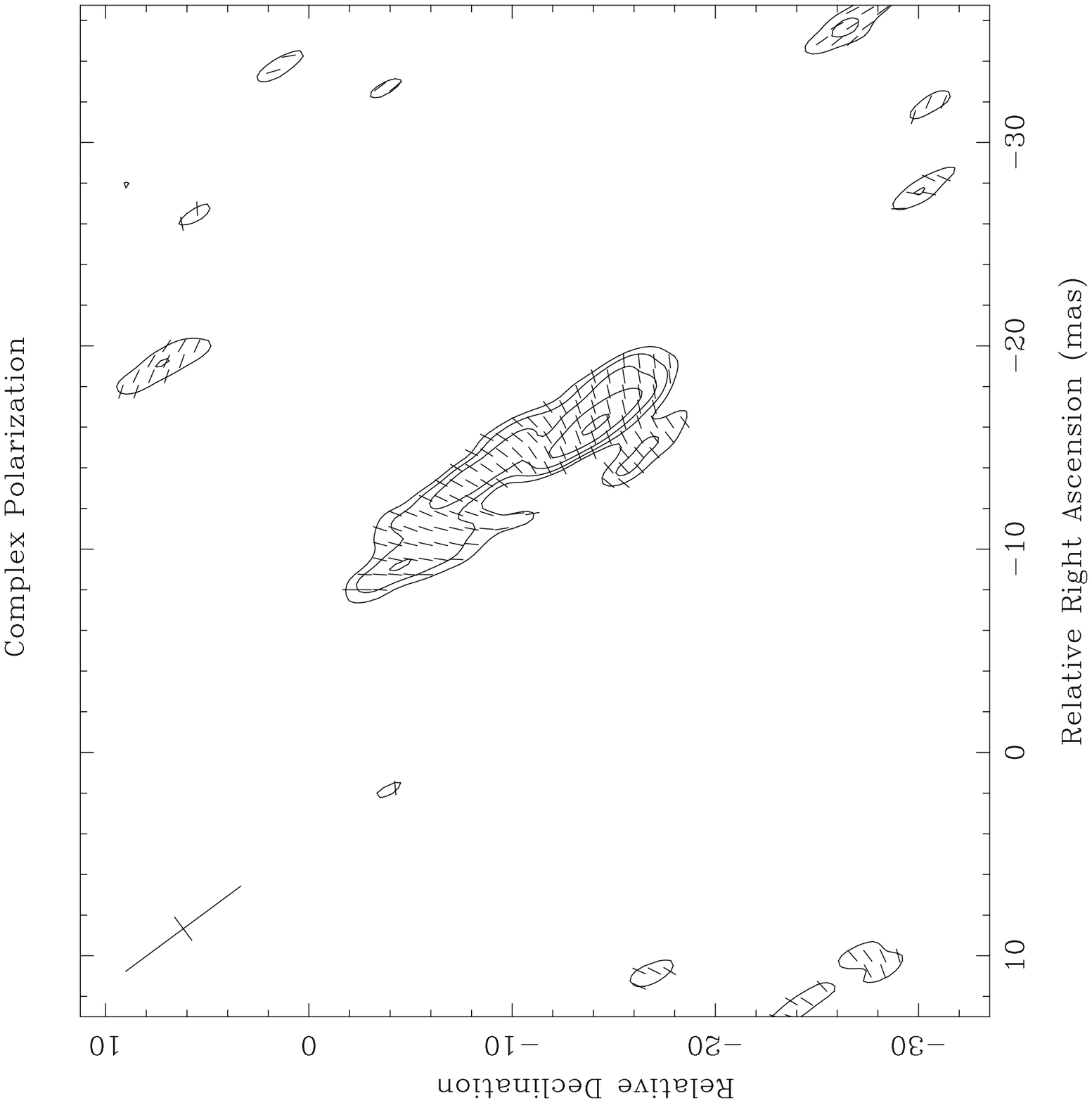,height=10.25cm}}
}
\vspace*{-0.75cm}

\hspace*{-0.7cm}
\mbox{
\rotate[r]{\psfig{file=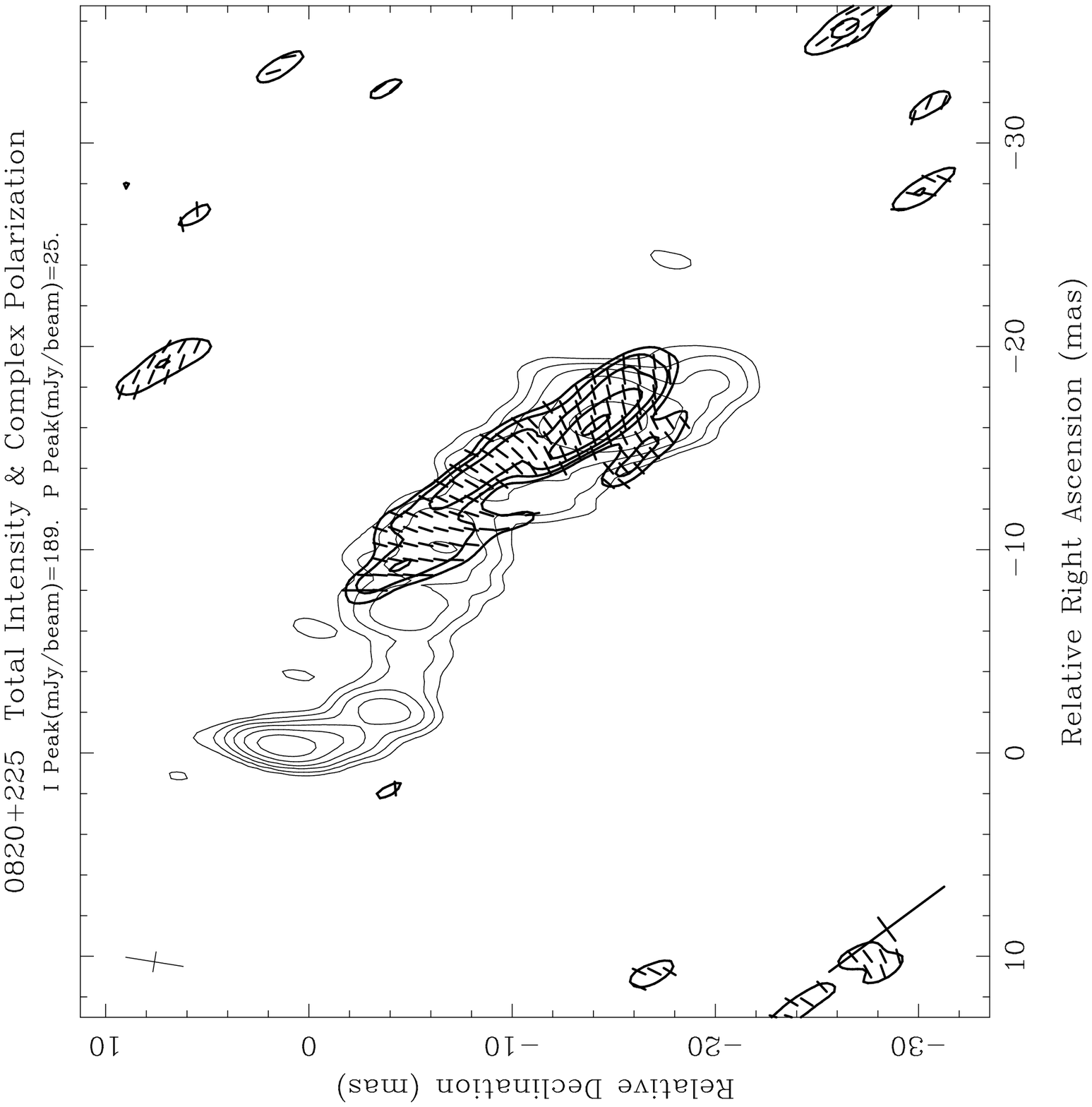,height=10.25cm}}
}
\vspace*{-0.75cm}

\caption{VLBI images of 0820+225 at 5~GHz at epoch 1995.36: (a) $I$ with
contours at $-1.8$, 1.8, 2.6, 3.6, 5.1, 7.2, 10, 14, 20,
29, 41, 58, and 82\% of the peak brightness of 188.7 mJy/beam. (b) $p$
with contours at 24, 34, 45, 64, and 90\% of the peak brightness of
25 mJy/beam and $\chi$
sticks superimposed. (c) The $I$ image from (a) with
every other contour omitted and $\chi$ sticks supermposed. Note that the
$I$ and $P$ beams are quite different in this case.}
\label{fig:gc16a}
\end{figure}

\begin{figure}
\mbox{
\psfig{file=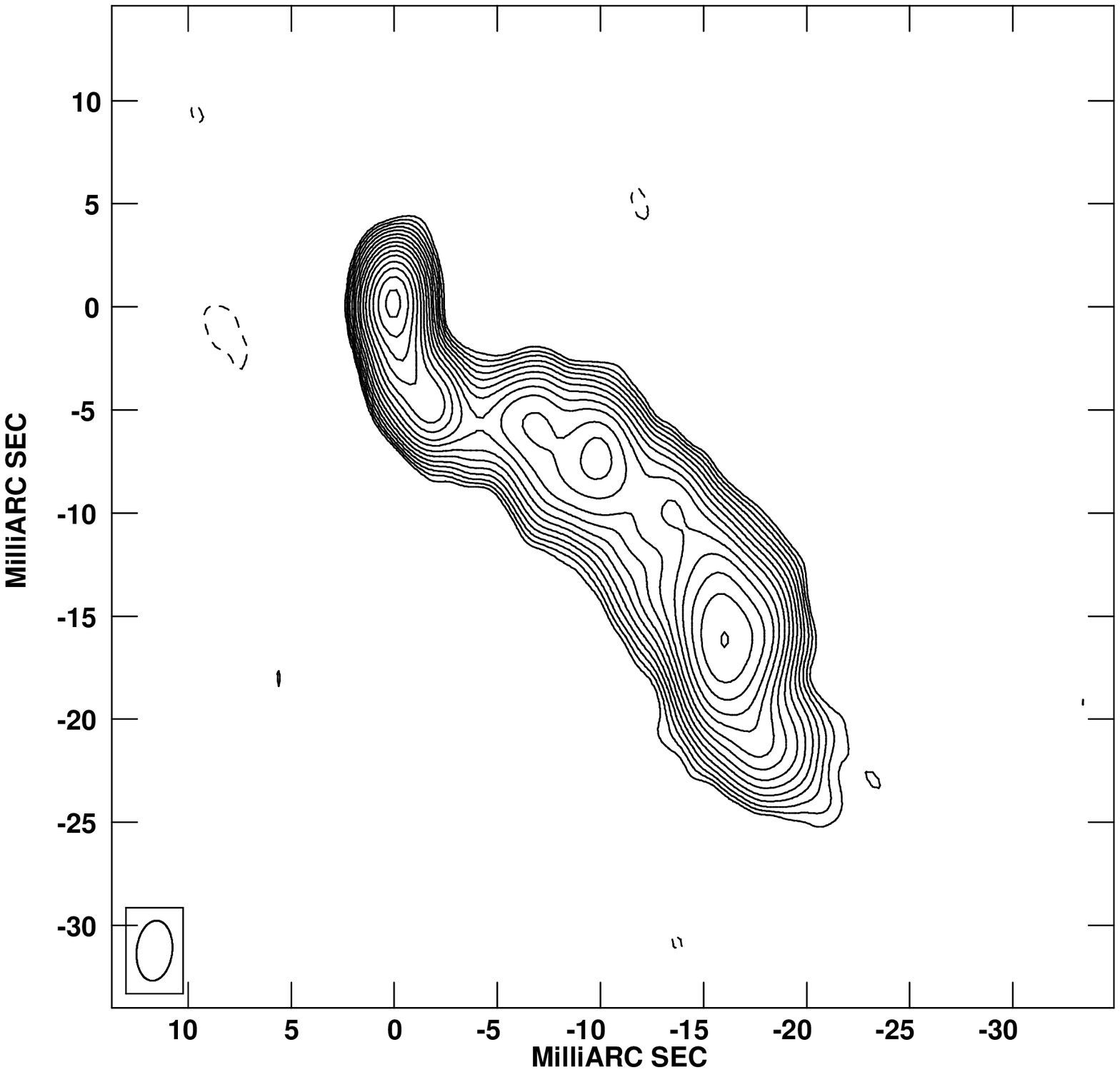,height=8.75cm}
}
\vspace*{-1.75cm}

\mbox{
\psfig{file=0820_6CM_P.PS,height=8.5cm}
}
\vspace*{-1.25cm}

\mbox{
\psfig{file=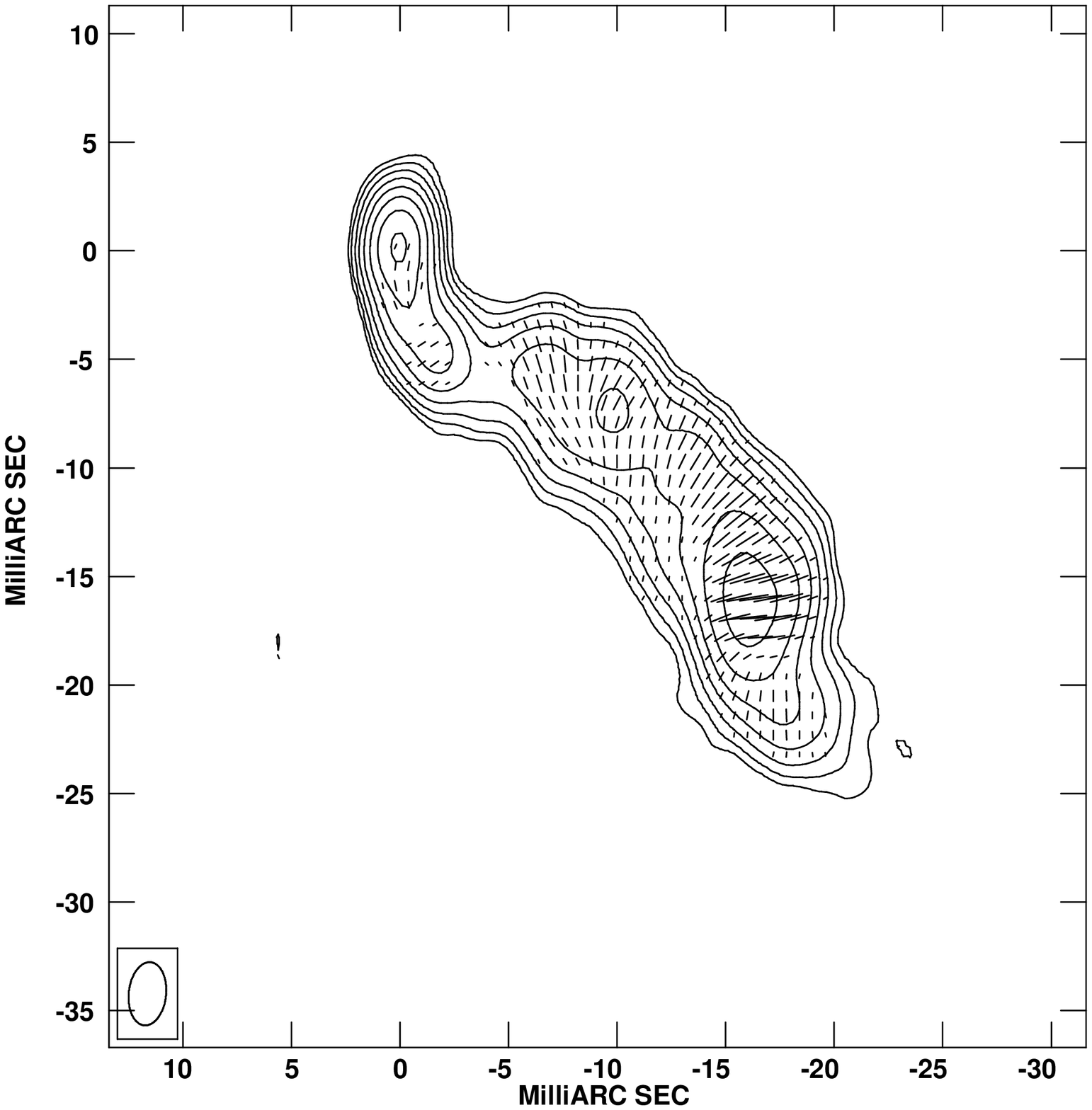,height=8.75cm}
}
\vspace*{-1.5cm}

\caption{VLBI images of 0820+225 at 5~GHz at epoch 1997.11: (a) $I$ with
contours at $-0.7$, 0.7, 1.0, 1.4, 2.0, 2.8, 4.0, 5.6, 8.0, 11, 16, 23,
32, 45, 64, and 90\% of the peak brightness of 167.1 mJy/beam. (b) $p$
with contours at 8, 11, 16, 23, 32,
45, 64, and 90\% of the peak brightness of 13.2 mJy/beam and $\chi$
sticks superimposed. (c) The $I$ image from (a) with
every other contour omitted and $\chi$ sticks supermposed.}
\label{fig:6cm}
\end{figure}

Figures~\ref{fig:gc16a}--\ref{fig:2cm} show contour images of
the total or linearly polarized flux density, with superposed sticks
showing the orientation of the polarization position angles $\chi$.
In each of the images, the restoring beam is indicated by a cross or
ellipse in some corner of the map. Note that the $I$ and $P$ beams for
the 1995.36 images in Fig.~\ref{fig:gc16a} are rather different; we
show the $P$ image restored with its own beam in order to accurately
reflect the different baseline coverages for the $I$ and $P$ datasets.

\subsection{Total Intensity Structure}

\begin{figure}
\mbox{
\psfig{file=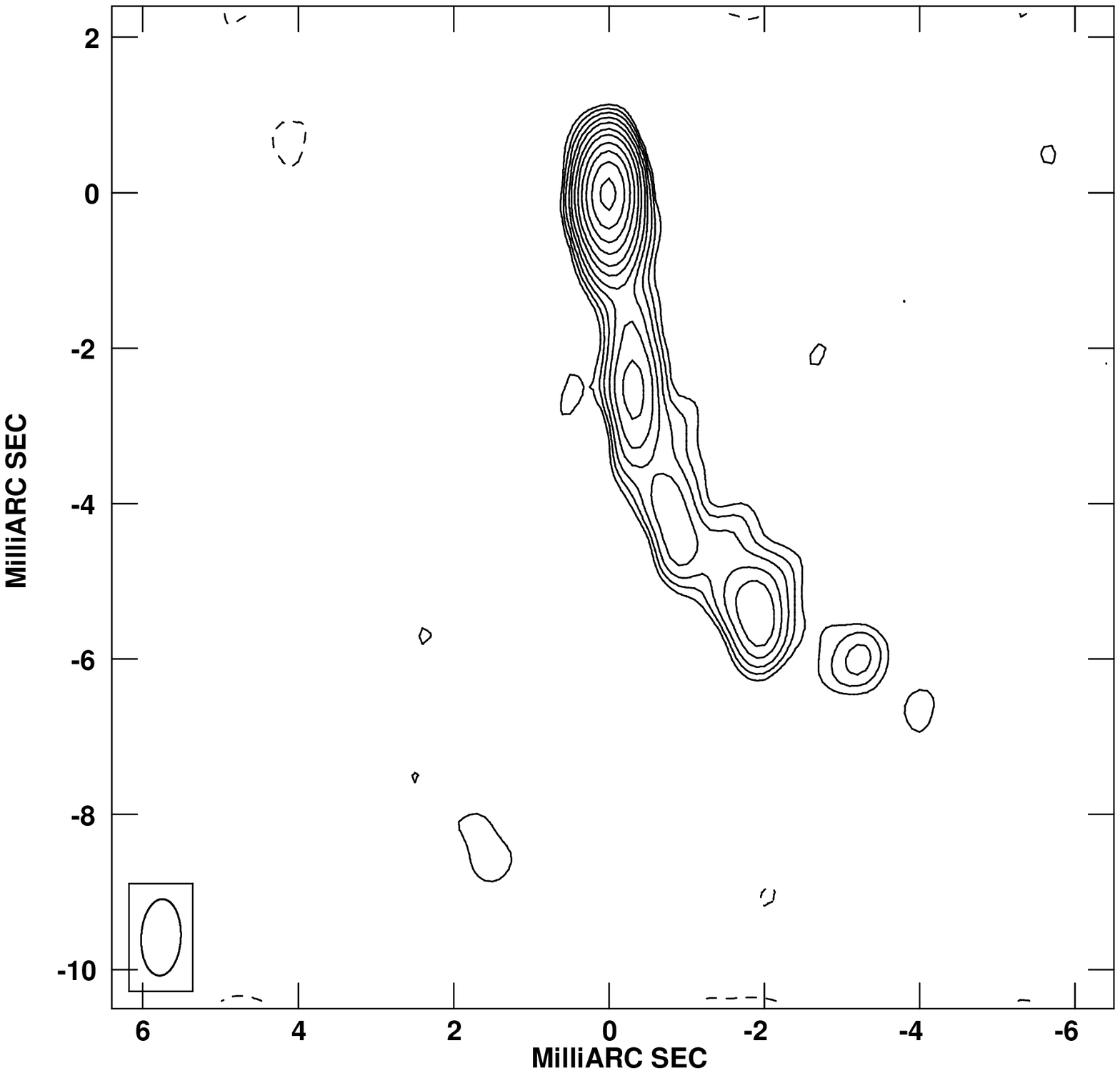,height=8.5cm}
}
\vspace*{-1.2cm}

\mbox{
\psfig{file=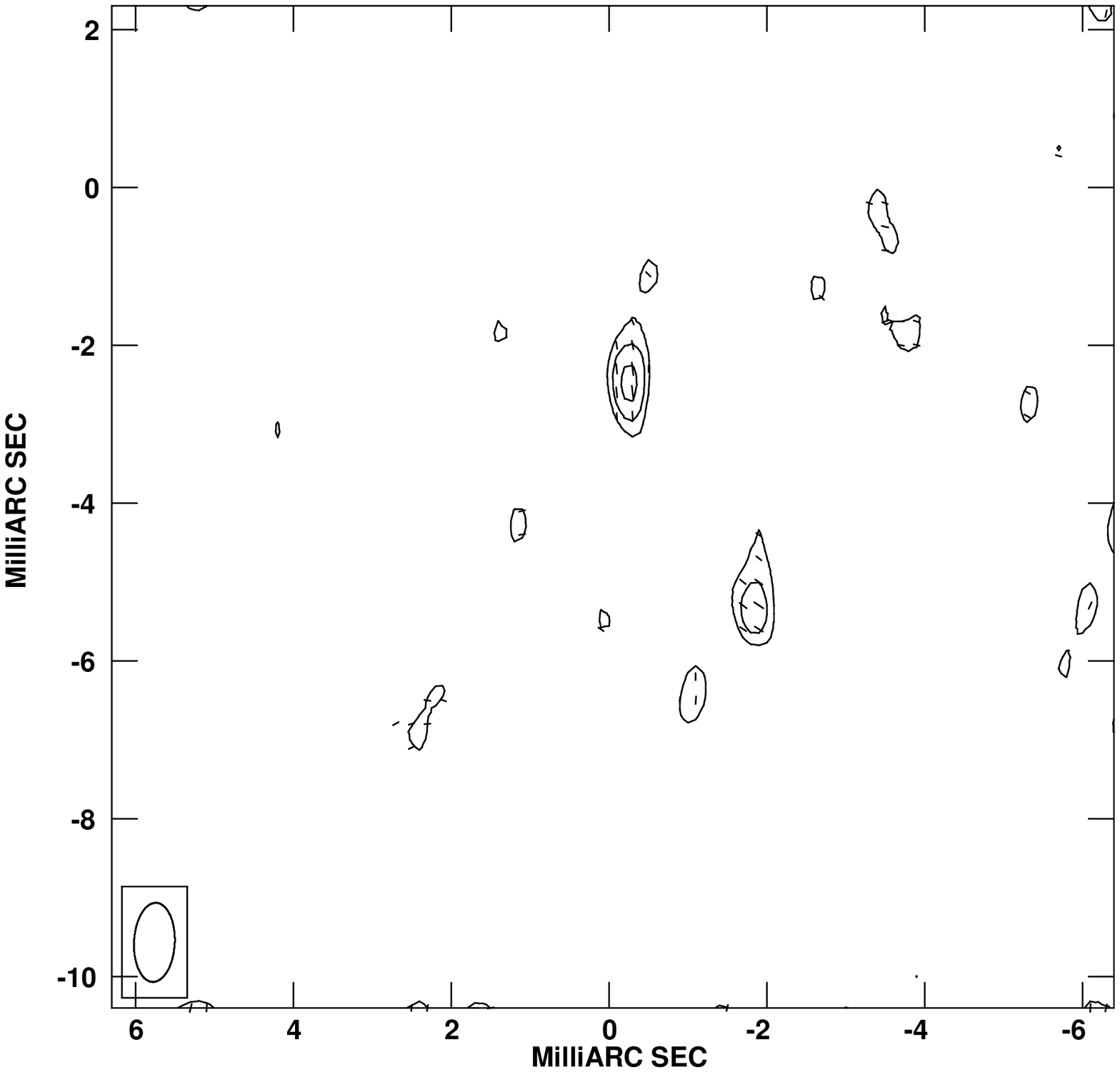,height=8.5cm}
}
\vspace*{-1.2cm}

\mbox{
\psfig{file=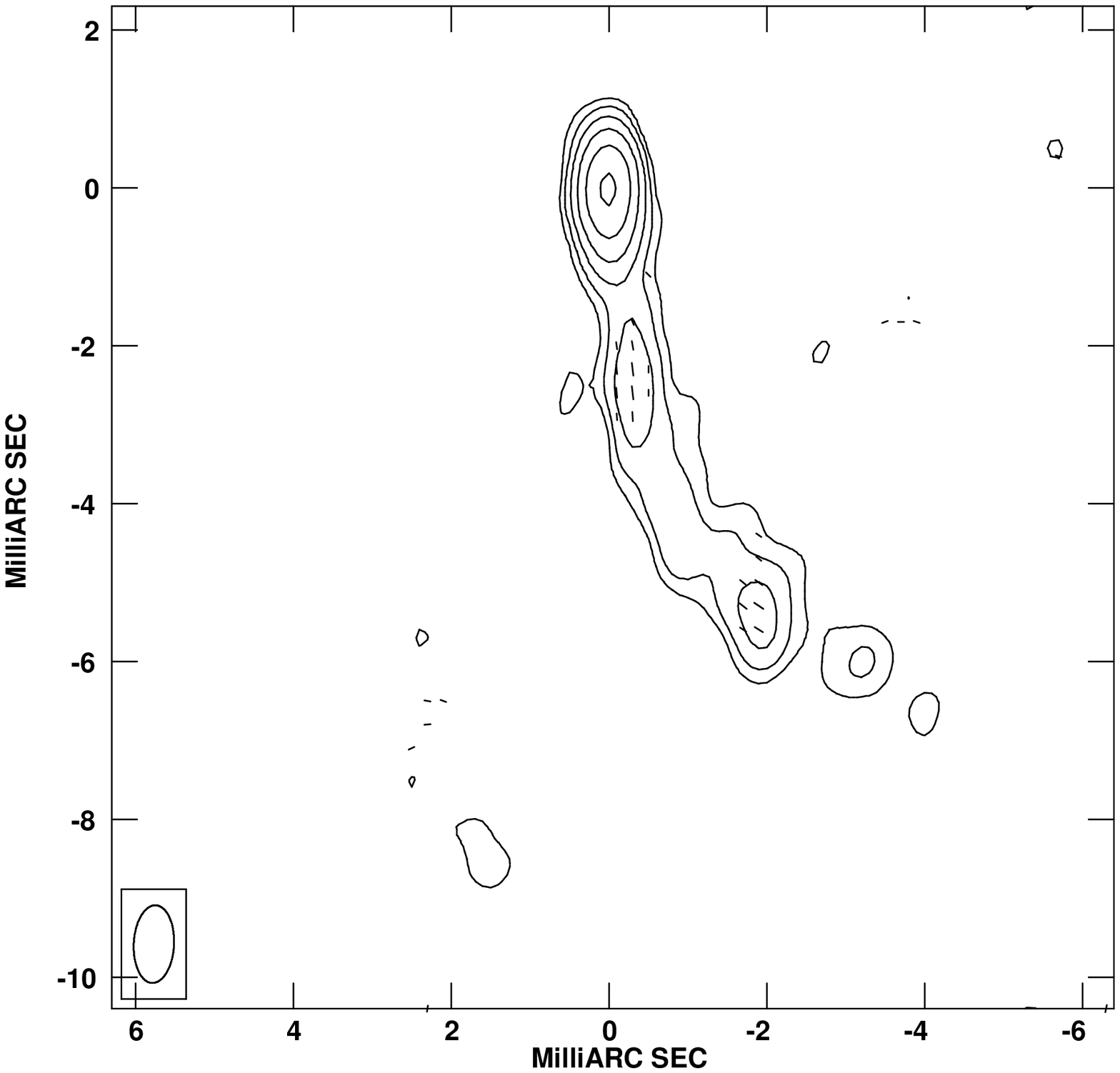,height=8.5cm}
}
\vspace*{-1.0cm}

\caption{VLBI images of 0820+225 at 15~GHz at epoch 1997.11: (a) $I$ with
contours at $-2.8$, 2.8, 4.0, 5.6, 8.0, 11, 16, 23,
32, 45, 64, and 90\% of the peak brightness of 109.0 mJy/beam. (b) $p$
with contours at
45, 64, and 90\% of the peak brightness of 3.4 mJy/beam and $\chi$
sticks superimposed. (c) The $I$ image from (a) with
every other contour omitted and $\chi$ sticks supermposed.}
\label{fig:2cm}
\end{figure}

The VLBI structure of 0820+225 is very rich in both $I$ and $P$. Our
5~GHz images, shown in Figs.~\ref{fig:gc16a} and \ref{fig:6cm},
follow the VLBI jet roughly 30~mas from the core. For ease of comparison,
the contour levels in mJy for the $I$ images in Fig.~\ref{fig:gc16a}
and the highest 12 contour levels in mJy for the $I$ images in 
Fig.~\ref{fig:6cm} coincide. We can see that, in contrast to the overwhelming
majority of BL Lacertae objects, the bulk of the parsec-scale flux is
contained in the jet rather than the core. The VLBI jet bends quite
strongly, in a shape that is suggestive of some type of precession or
oscillatory instability.
The 8.4~GHz image in Fig.~\ref{fig:4cm}a shows the extended VLBI jet in
somewhat more detail. The jet gives the impression of being continuous,
but also shows clear clumpiness in the brightness distribution.
The 15~GHz image in Fig.~\ref{fig:2cm}a shows the inner part of the curved
VLBI jet with higher resolution. The knotty structure of the jet is
clearly visible.

\begin{figure}
\hspace*{1.0cm}
\mbox{
\psfig{file=0820_4CM_I.PS,height=8.5cm}
}
\vspace*{-1.5cm}

\hspace*{1.0cm}
\mbox{
\psfig{file=0820_4CM_P.PS,height=8.5cm}
}
\vspace*{-1.5cm}

\hspace*{1.0cm}
\mbox{
\psfig{file=0820_4CM_2P.PS,height=8.5cm}
}
\vspace*{-1.5cm}

\caption{VLBI images of 0820+225 at 8.4~GHz at epoch 1997.11: (a) $I$ with
contours at $-1.4$, 1.4, 2.0, 2.8, 4.0, 5.6, 8.0, 11, 16, 23,
32, 45, 64, and 90\% of the peak brightness of 112.2 mJy/beam. (b) $p$
with contours at 23, 32,
45, 64, and 90\% of the peak brightness of 4.5 mJy/beam and $\chi$
sticks superimposed. (c) The $I$ image from (a) with
every other contour omitted and $\chi$ sticks supermposed.}
\label{fig:4cm}
\end{figure}

Figure~\ref{fig:sidist} shows the distribution of the spectral index
calculated between 5 and 8.4 GHz, $\alpha_{5-8}$ ($S\propto\nu^{\alpha}$).
We can see that all regions of the VLBI structure have steep spectra,
indicating that they are optically thin. The spectral
index throughout most of the outer jet is $\alpha_{5-8}\simeq -2.0$ to
$-2.5$, while the spectra are less steep near the core and in the
inner jet ($\alpha_{5-8}\simeq 0$ to $-1$). The spectrum is actually
flattest near the westward bend, near the position of
C2 (see Fig.~\ref{fig:rmdist}). 

\subsection{Rotation Measure Distribution on Parsec Scales}

\begin{figure}
\mbox{
\rotate[r]{\psfig{file=0820_RMPLOT.PS,height=8.0cm}}
}
\mbox{
\psfig{file=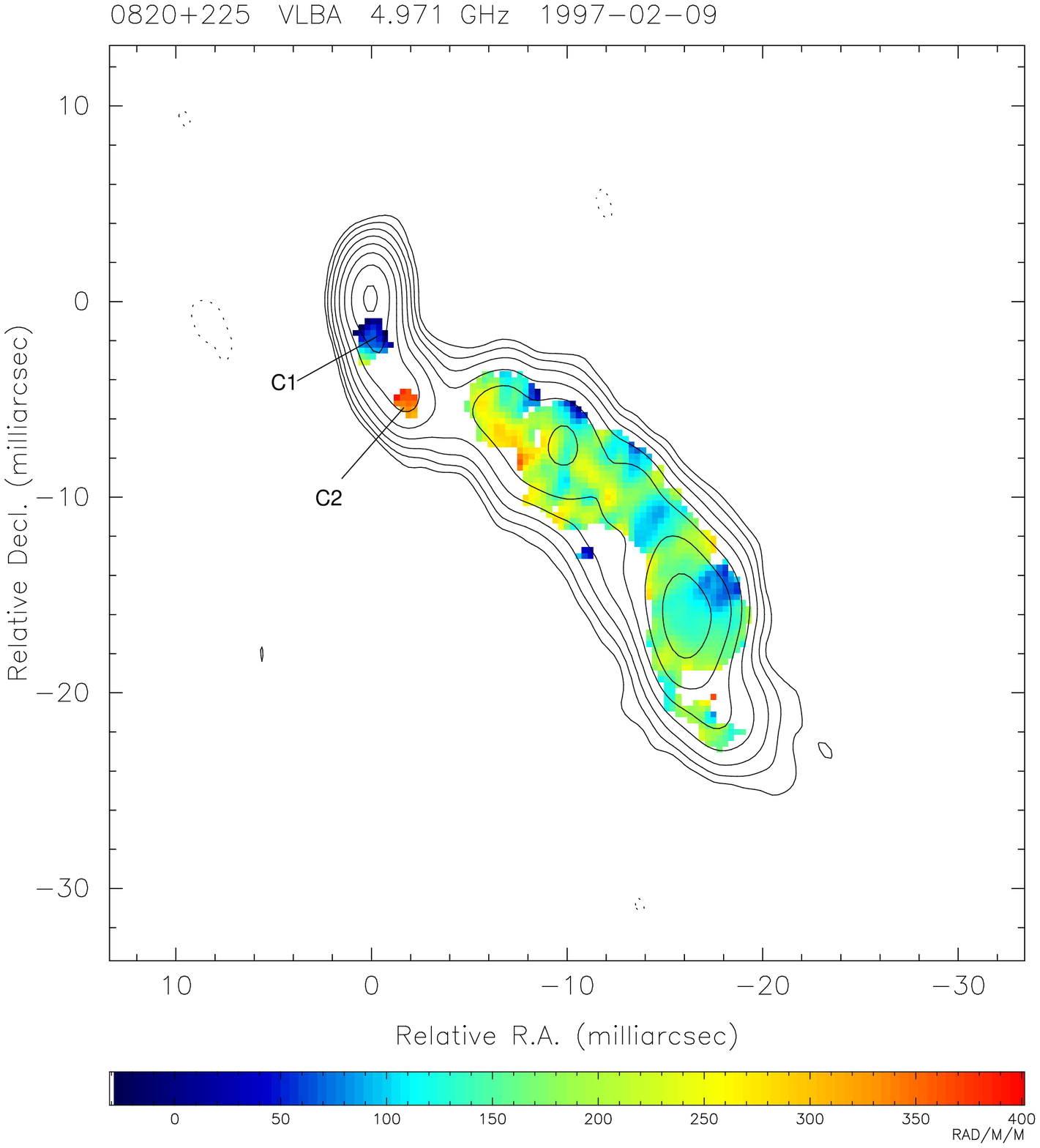,height=10.5cm}
}
\caption{Top: Polarization position angles $\chi$ as a function of
the square of the wavelength $\lambda^2$ for the two inner
knots in the VLBI jet of 0820+225. The wavelength dependence of
both components is consistent with a $\lambda^2$ law, but the
rotation measure of the second knot, ``C2,'' is clearly much higher
than that of the knot closest to the core, ``C1.'' Bottom: $I$ image of 
0820+225 at 5~GHz from Fig.~\ref{fig:6cm}c with the rotation-measure 
distribution calculated from the 5 and 8.4-GHz polarization maps 
superimposed.}
\label{fig:rmdist}
\end{figure}

Only two polarized features are detected in the inner part of the
jet, and these are detected at all three frequencies.
A comparison of the inner-jet region in the polarization images in
Figs.~\ref{fig:6cm}b, \ref{fig:4cm}b, and \ref{fig:2cm}b immediately
demonstrates the presence of a non-uniform frequency dependence for
the observed polarization position angles. The knot roughly 3~mas
from the core (``C1'') has nearly the same $\chi$ values at all three
frequencies; in
contrast, the three $\chi$ values for the knot roughly 6~mas from
the core (``C2'') are very different. The top plot in
Fig.~\ref{fig:rmdist} clearly shows that the observed $\chi$ values
for both knots are consistent with $\lambda^2$ dependences ($\lambda = c/\nu$
is the observing wavelength),  as expected for Faraday rotation: the 
inferred rotation measure for the inner knot is only +42~rad/m$^2$, while 
that for the outer knot is +353~rad/m$^2$.

Polarized flux further from the core is detected only at 5 and 8.4~GHz.
Therefore, we cannot strictly test there for a $\lambda^2$ dependence
of the measured $\chi$ values. Nonetheless, we can compare the
$\chi$'s observed at the two frequencies and calculate the rotation
measure implied if any difference is, indeed, due to external Faraday
rotation. Figure~\ref{fig:rmdist} (bottom) shows the distribution of the
inferred RMs based on a comparison of the images for the two lower
frequencies; in the inner jet, these RMs are consistent with those
derived using all three frequencies. There is evidence
for inhomogeneity in the RM distribution in the outer jet
as well; there is another region of high RM just to the west of the
dip in the jet intensity about 8~mas from the core.

\begin{figure}
\mbox{
\psfig{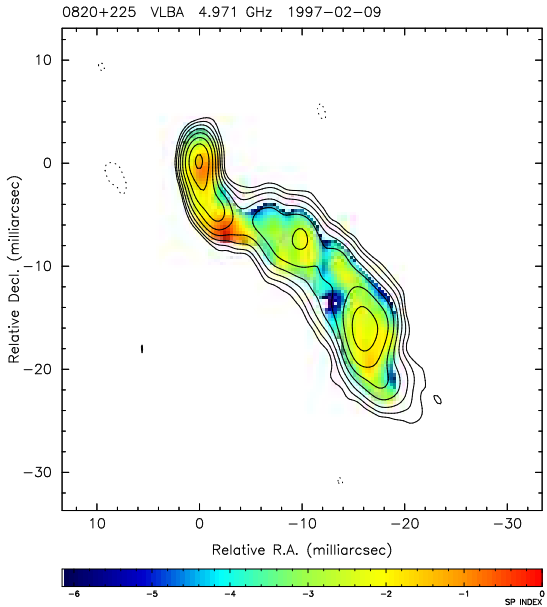}
}
\vspace*{-0.3cm}
\caption{$I$ images of 0820+225 at 5~GHz at epoch 1997.11 from
Fig.~\ref{fig:6cm}c with the distribution of the spectral index
calculated between 5 and 8.4 GHz supermposed.}
\label{fig:sidist}
\end{figure}

In general, the observed RMs decrease with distance from the core in
the outer part of the jet. The RMs in the outer jet remain somewhat 
inhomogeneous, but have more modest values, some roughly comparable to 
the integrated RM determined using 1.50 and 1.67~GHz VLA observations, 
$+81$~rad/m$^2$ (Pushkarev 2001). Since the 8.4 and 5~GHz polarization 
images presented
here are already dominated by the outer jet, and this trend should
continue toward lower frequencies, we expect that the 1.50--1.67~GHz 
polarization will be dominated by the contribution of regions lying
$\simeq 15$~mas or more from the core. The fact that the RMs inferred
here for some parts of the outer jet are roughly comparable to the integrated 
value may indicate that we are just starting to detect a constant 
``foreground,'' presumably Galactic, rotation measure. However, a
substantial fraction of the outer 5-GHz jet has RMs that appreciably
exceed the nominal integrated value, so that there is probably some
contribution to the observed RMs from thermal plasma in the immediate
vicinity of the BL Lac object throughout most of the observed 5-GHz
VLBI jet. There is also some evidence that the observed RMs tend to
be higher toward the southern edge of the jet; the physical origin for 
such a gradient is not clear.

Comparing Figs.~\ref{fig:sidist} and \ref{fig:rmdist} (bottom), we can 
see that
the 5--8.4~GHz spectrum flattens in the core (as is expected) and also
near the westward bend in the jet, where the rotation measures are highest.
This is suggestive of either low-frequency absorption near this bend,
or possibly of re-acceleration of electrons at this location. Both would
be consistent with an interaction between the jet and external medium,
and it is possible that this is related to the bending of the jet itself.
We cannot test this idea more fully, however, since we were not able to
image the region near and beyond this bend at 15~GHz, where we would
expect absorption to be appreciably lower than at 5~GHz.

\subsection{Linear Polarisation Structure After ``De-rotation''}

\begin{figure}
\mbox{
\psfig{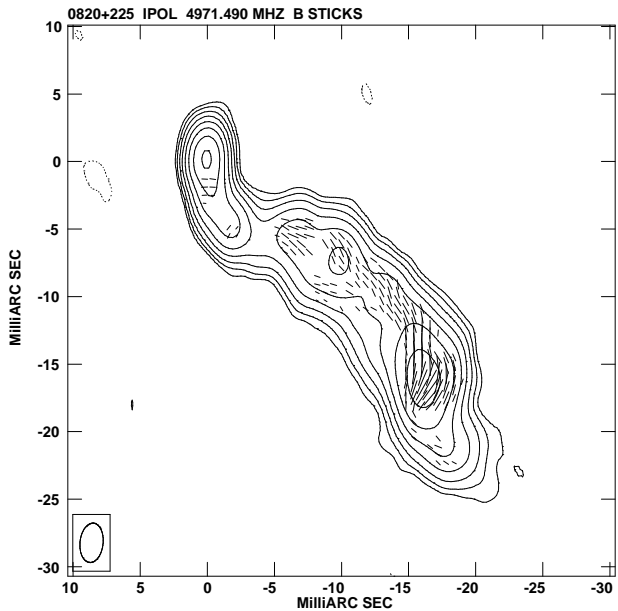}
}
\caption{$I$ image of 0820+225 at 5~GHz from Fig.~\ref{fig:6cm}c
with the inferred intrinsic underlying {\bf B} field superimposed.}
\label{fig:bdist}
\end{figure}

Thus, the rotation-measure distribution is quite non-uniform, and
the Faraday rotations implied in the regions of highest rotation
measure at the lowest observed frequency can reach $\sim 60-70^{\circ}$.
Therefore, it is essential to correct for the observed rotation
measure distribution if we wish to infer the intrinsic values for 
$\chi$, and thereby the underlying magnetic field structure.
Figure~\ref{fig:bdist} shows the result of this correction. Here,
we have adjusted the observed $\chi$ values in accordance with the
rotation-measure distribution in Fig.~\ref{fig:rmdist} (bottom), then rotated
the result by $90^{\circ}$ so that it shows the distribution of the
underlying magnetic field implied by the ``de-rotated'' $\chi$
values. Recall that all the observed 5~GHz components, including the
core, are dominated by optically thin emission (Fig.~\ref{fig:sidist}),
so that {\bf B} is perpendicular to the polarization electric
vector.

This figure shows that
the magnetic field in the inner part of the jet, before the westward
bend, is transverse to the local jet direction, and that the dominant
field becomes longitudinal beyond this bend. In addition, inspection
of Figs.~\ref{fig:4cm}b and \ref{fig:6cm}b shows that the polarized
emission in the outer part of the VLBI jet is offset toward the northern
edge of the main ridge lines in the corresponding $I$ images. These
properties are suggestive of an interaction between the jet and
surrounding medium that enhances the longitudinal field component
due to shear. The dominant magnetic field is longitudinal essentially
throughout the outer portion of the jet. The inferred magnetic field
pattern in the bright component roughly 20~mas from the core is
complex. If the magnetic field sticks in this component in
Fig.~\ref{fig:bdist} reflect the direction of the underlying jet
flow, this indicates that the jet turns sharply southward, then back
eastward before continuing westward at the end of the visible jet; this 
is plausible, but must be verified by analysis of component motions 
in this region.

\begin{figure}
\mbox{
\psfig{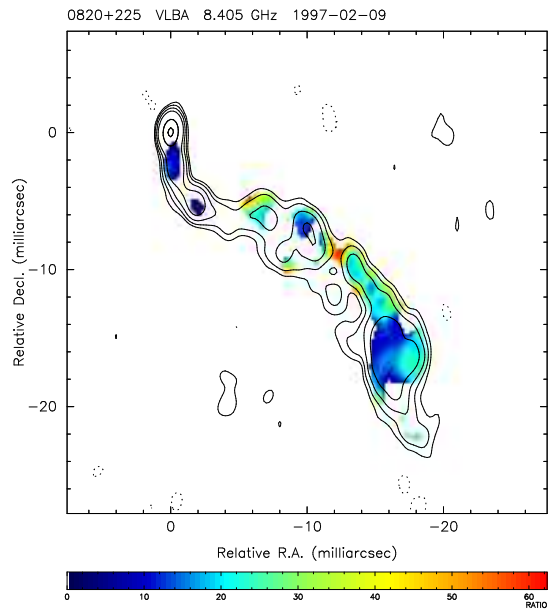}
}
\mbox{
\psfig{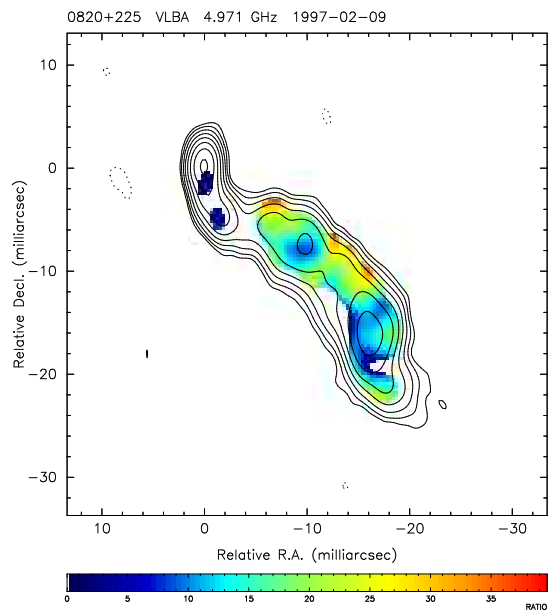}
}
\caption{8.4 and 5~GHz $I$ images from (top) Fig.~\ref{fig:4cm}b and 
(bottom) Fig.~\ref{fig:6cm}b with the corresponding
distributions of degree of polarization $m$ displayed in grey scale
superimposed.}
\label{fig:moni_4+6cm}
\end{figure}

Figure~\ref{fig:moni_4+6cm} shows distributions 
of the degree of polarization
$m$ superposed on $I$ images at 8.4 and 5 GHz. In each case, we
include only those $m$ values for pixels where both polarization and
total-intensity were clearly detected. We can see 
that $m$ in the jet is typically $\simeq 10-15\%$, and reaches 
30--40\% or more in some places. In the 5~GHz $m$ distribution,
the degree of polarization increases toward the outer edge of the curved jet
structure, consistent with the possibility that the longitudinal
field has been more highly ordered there, where the shear interaction
is strongest. The most unusual feature in the $m$ distribution at 8.4 GHz
is the rather high inferred polarization in a region of low total
intensity about 13~mas west and 8~mas south of the core. The map in
Fig.~\ref{fig:4cm}b clearly shows the detection of polarized flux in
this location, and gives no indication of a dip in polarized flux in
the location of the dip in total intensity. If this region of high
polarization -- indicating a region of very highly ordered longitudinal
field -- is real, its origin is not obvious. One possibility is that
the regions occupied by various components in the extended part of the
jet have a much stronger transverse field component than this low-intensity
region, which partially cancels the longitudinal field due to shear.

\subsection{Estimation of the Magnetic-Field Strength and Properties
of the Thermal Gas}

Let us suppose that the thermal gas giving rise to the large RM in
the vicinity of the westward bend just beyond ``C2'' has
properties fairly typical of a narrow-line cloud. In particular,
we will assume that its temperature is $T\simeq 10^4$~K, as has been
established from observations of forbidden and semi-forbidden lines
(Koski 1978; Heckman \& Balick 1979). If we then assume further that 
this cloud is in equipartition, we can derive at least a rough estimate 
for the magnetic field in this region using the relations (Burn 1966;
Tucker 1975)
$$
{\rm RM} = 8.1\times 10^5 \int N_e\vec{B}\cdot d\vec{l}
$$
$$
{\rm RM} \simeq8.1\times 10^5 N_eB_\parallel L\,
$$
$$
N_ekT \simeq \frac{B^2}{8\pi}
$$
where the first equation gives the definition of the rotation measure 
and the second is an approximation to the RM for the case of electromagnetic
radiation travelling a length $L$ through a region of homogeneous 
thermal plasma with the magnetic-field component along the line of 
sight equal to $B_\parallel$.  Here, $L$ is in pc, $B$ and $B_\parallel$ 
are in G, the electron number density $N_e$ is in cm$^{-3}$, and RM is in 
rad\,m$^{-2}$. Eliminating $N_e$ and multiplying the observed rotation 
measure by $(1+z)^2$ to determine the RM in the source frame,  we obtain
$$
B^2B_\parallel \simeq \frac{RM(1+z)^2\pi kT}{10^5L}.
$$
Approximating $B_\parallel\simeq B$, as we expect if the magnetic
field has no special orientation relative to the line of sight to
the radio source (this approximation should affect the final result by at
most about a factor of two or so),
$$
B^3 \simeq \frac{{\rm RM}(1+z)^2\pi kT}{10^5L}
$$
$$
B \simeq \left[\frac{{\rm RM}(1+z)^2\pi kT}{10^5L}\right]^{1/3}
$$
$$
B \simeq 3.8\times 10^{-5}L^{-1/3}.
$$
\noindent
Further, we have some idea of the size of the region of thermal
gas giving rise to the rotation measure, based on the extent of the
region of high rotation measure on the sky -- say, about 5~mas.
If the thermal-gas cloud were much bigger than this,
we would expect the observed region of high rotation measure to be larger.
Since 1~mas = 5.64~pc at the redshift of 0820+225 ($H_o = 
75$\,km\,s$^{-1}$\,Mpc$^{-1}$, $q_o = 0.5$), this implies a diameter 
$L$ of about 20--30~pc, which is reasonable, given estimates of the 
size of the narrow-line region for nearby AGNs ($r\simeq 100$~pc or 
slightly larger; Peterson 1997). Substituting this value into our 
expression for $B$, we obtain the estimate $B\simeq 12-14 \mu$G for 
this region several tens of parsecs from the VLBI core, consistent 
with general expectations for the strength of the magnetic field on 
parsec scales (perhaps $10-100 \mu$G) and the inferred magnetic field 
on kiloparsec scales ($\simeq 0.15-1.5 \mu$G; see for example, 
Laing \& Bridle 1987). While it does not provide us with precise 
estimates, this analysis does suggest parameters for the thermal gas 
associated with the Faraday rotation and the region in which it is located 
that are consistent with current concepts about narrow-line clouds.

\section{Conclusion}

Based on our earlier 5~GHz $I$ and $P$ VLBI images of 0820+225
(Gabuzda, Pushkarev \& Cawthorne 2000), we already knew that this was 
an unusual source.
It made it into the K\"uhr \& Schmidt (1990) 1-Jy sample of BL Lac
objects because its integrated flux at 5 GHz is, indeed, more than
1 Jy. Like the other sample sources, a high fraction of its integrated
flux is on milliarcsecond scales. However, in sharp contrast to
the other sources in the sample, only a small fraction of the flux
on VLBI scales is contained in the VLBI core. The jet is very extended,
especially given that the redshift of the source is relatively large
for a BL Lac object, $z = 0.95$; this means that the VLBI jet sprawls
over some 150--200 pc, with much of the emission located at substantial
distances from the VLBI core.

The multi-frequency VLBA images presented here have underlined the
unusual nature of this object by revealing an extremely non-uniform
distribution for the rotation-measure on milliarcsecond scales.
Galactic clouds of thermal plasma would be far too large to account
for inhomogeneity in the rotation measure on such small angular scales.
It is virtually certain that the thermal plasma giving rise to the
non-uniform Faraday rotation is in the immediate environment of the
VLBI jet. Therefore, since the RM is larger by a factor of $(1+z)^2$
in the rest frame of the source, the intrinsic RM of the component
``C2'' 6~mas from the core is about 1300\,rad\,m$^{-2}$, and the RM gradient
in this part of the jet is $\simeq 400$\,rad\,m$^{-2}$\,mas$^{-1}$.

%

%
%

In addition, we have found compelling evidence for the formation of
a shear layer in the outer part of the VLBI jet of 0820+225. While
the inferred magnetic field in the inner part of the jet (after correction
for the measured RM distribution) is transverse, the predominant
magnetic field in the outer jet becomes longitudinal. The ordering
of the longitudinal field in the outer jet is substantial, as indicated
by the observed degrees of polarization at both 5 and 8.4~GHz, which
reach $\simeq 30-40\%$.  The degree of polarization across
the 5-GHz jet increases toward the outer edge, consistent with the idea
that it is here that the enhancement of the longitudinal field component
by the shear interaction is strongest, as might be expected.

Thus, 0820+225 provides us with an interesting, seemingly paradoxical,
combination of properties: very weak optical line emission, but clear
evidence for the presence of appreciable quantitites of thermal gas in
the immediate vicinity of the AGN. This suggests that the weak line
emission in this object is not simply due to a lack of gas: the gas
is present, but is either not exposed to ionizing continuum radiation
or is in a hot phase not conducive to the formation of dense
clouds (Corbett et al. 1996).

Given the recent detection of a non-uniform parsec-scale RM distribution
in BL Lac itself as well (Reynolds et al. 2001), this  presents the
interesting possibility that the weak optical line emission characteristic
of these sources is generally not associated with a dearth of thermal gas.
Since there is no clear evidence that BL Lac objects are more beamed
than core-dominated quasars (Ghisellini et al. 1993), it is also implausible
that the optical continua of BL Lac objects as a class are so much
more highly beamed than the continua of quasars that their emission lines
are ``swamped,'' giving rise to a featureless continuum. Thus, the most
simple and plausible explanation for the weakness of the optical emission
lines of BL Lac objects may be that the ionizing continuum radiation is
insufficient to produce optical lines as luminous as those in quasars.
In the framework of unification schemes identifying the FR~I radio
galaxies as the parent population of BL Lac objects (Browne 1983; Wardle,
Moore \& Angel 1984), this is also consistent with the analysis of
Baum et al. (1995), who found the optical emission lines of FR~I
radio galaxies to be systematically less luminous than those of FR~II
radio galaxies of the same total radio luminosity or radio core
power, and concluded that this was most likely associated with a lack
of ionizing uv continuum in FR~I sources, rather than a lack of thermal gas.

We are currently engaged in a multi-epoch, multi-frequency VLBA project
to study this object in more detail, and to search for possible
superluminal motions and variations in the RM distribution. Such variations
do not require changes in the Faraday screen (movement of blobs of
thermal plasma, for example); the observed RM distribution could vary
due to changes in the distribution of polarized emission behind the
non-uniform Faraday screen. In this way, we hope that our multi-epoch
study will provide further valuable information about both the
distribution of thermal plasma in the vicinity of the BL Lac object
and the properties of the unusual VLBI jet itself.

\section{Acknowledgements}

ABP acknowledges support from the Russian Foundation for Basic
Research (project code 99-0217799). DCG acknowledges support from the 
European Commission under the IHP Programme (ARI) contract No. 
HPRI-CT-1999-00045.  We thank L. Gurvits for useful discussions in
connection with this work.

\clearpage

\end{document}